\documentclass[a4paper,11pt]{article}
\pdfoutput=1 

\usepackage{jheppub} 

\usepackage[T1]{fontenc} 

\usepackage[dvipsnames]{xcolor}

\def\figureautorefname~#1\null{Fig.\,#1\null}

\def\equationautorefname~#1\null{Eq.\,(#1)\null}

\newcommand{\inab}{\,{\rm ab}^{-1}}

\newcommand{\A}{\mathcal{A}}

\newcommand{\bpm}{\begin{pmatrix}}
\newcommand{\epm}{\end{pmatrix}}

\newcommand{\la}{\langle}
\newcommand{\ra}{\rangle}

\newcommand{\aall}{\gamma\gamma\to \ell^+\ell^-}
\newcommand{\aaqq}{\gamma\gamma\to q\bar{q}}
\newcommand{\aqaq}{\gamma q\to \gamma q}
\newcommand{\abab}{\gamma b\to \gamma b}

\newcommand{\pllp}{pp \to p\,\ell^+\ell^-\,p}

\newcommand{\eeaa}{e^+e^- \to \gamma \gamma}
\newcommand{\mumuaa}{\mu^+\mu^- \to \gamma \gamma}

\newcommand{\eemmaa}{e^+e^- (\mu^+\mu^-) \to \gamma \gamma}

\title{Probing positivity at the LHC with exclusive photon-fusion processes}

\author[a,b]{Jiayin Gu,}
\author[a,c]{Chi Shu}

\affiliation[a]{Department of Physics and Center for Field Theory and Particle Physics, \\ Fudan University, Shanghai 200438, China}
\affiliation[b]{ Key Laboratory of Nuclear Physics and Ion-beam Application (MOE), \\ Fudan University, Shanghai 200433, China}
\affiliation[c]{Department of Physics and Enrico Fermi Institute, University of Chicago, Chicago, IL 60637}

\emailAdd{jiayin\_gu@fudan.edu.cn} 
\emailAdd{cshu20@fudan.edu.cn}

\abstract{By tagging one or two intact protons in the forward direction, it is possible to select and measure exclusive photon-fusion processes at the LHC.  The same processes can also be measured in heavy ion collisions, and are often denoted as ultraperipheral collisions (UPC) processes.  Such measurements open up the possibility of probing certain dimension-8 operators and their positivity bounds at the LHC.  As a demonstration, we perform a phenomenological study on the $\aall$ processes, and find out that the measurements of this process at the HL-LHC provide reaches on a set of dimension-8 operator coefficients that are comparable to the ones at future lepton colliders.  We also point out that the $\aqaq$ process could potentially have better reaches on similar types of operators due to its larger cross section, but a more detailed experimental study is need to estimate the signal and background rates of this process.  The validity of effective field theory (EFT) and the robustness of the positivity interpretation are also discussed.}

\begin{document} 
\maketitle
\flushbottom


\section{Introduction}

Probing dimension-8 (dim-8) operators of the Standard Model Effective Field Theory (SMEFT) at the Large Hadron Collider (LHC) is both an interesting and a challenging task.  
Certain dim-8 operators are subject to the so-called positivity bounds, assuming the underlying UV physics is consistent with the fundamental principles of quantum field theory (QFT), including unitarity, locality, analyticity and Lorentz invariance~\cite{Adams:2006sv,  Bellazzini:2016xrt, Bellazzini:2017bkb, 
deRham:2017avq, deRham:2017zjm, deRham:2018qqo,
Bellazzini:2018paj, Bi:2019phv, Remmen:2019cyz, Remmen:2020vts, Zhang:2020jyn,
Fuks:2020ujk, Yamashita:2020gtt, Gu:2020ldn,
Trott:2020ebl, Bonnefoy:2020yee, Chala:2021wpj, Zhang:2021eeo, Li:2022tcz, deRham:2022hpx, Li:2022rag, Hamoudou:2022tdn, Li:2022aby, Chala:2023jyx, Davighi:2023acq, Chen:2023bhu, Chala:2023xjy}.  Such positivity bounds do not exist for dim-6 operators without additional assumptions~\cite{Low:2009di, Falkowski:2012vh, Bellazzini:2014waa, Gu:2020thj,
Azatov:2021ygj, Remmen:2020uze, Remmen:2022orj, Davighi:2021osh, Altmannshofer:2023bfk}.  On the other hand, dim-6 operators are generally expected to provide the leading new physics contributions, assuming the scale of new physics is sufficiently higher than the energy scale of the collider so that the EFT expansion is valid in the first place.  This creates a conundrum for probing dim-8 operators and their positivity bounds at the LHC.  On the one hand, due to their energy enhancements ($\sim E^4$) we want to make use of the high energy events to maximize the sensitivity and to discriminate them from the dim-6 effects; on the other hand, high energy events typically have low statistics, and are poorly measured.  This leads to issues in the EFT interpretation, as the reach on the new physics scale $\Lambda$ tend to be comparable or even lower than the center-of-mass energy of the events. 
Furthermore, in these cases the dim-8-squared contribution can also be sizable.  
Probing positivity bounds requires us to measure the interference term between SM and the dim-8 contribution, which is sensitive to the signs of the Wilson coefficients.  While a sizable dim-8-squared contribution can be considered as an important signal for new physics (putting aside the EFT validity problem), it is effectively a background for probing positivity.  
Several studies pointed out the advantages of future lepton colliders in probing the dim-8 operators and positivity bounds~\cite{Fuks:2020ujk, Gu:2020ldn, Ellis:2019zex,Ellis:2020ljj}.\footnote{For comparisons with the reaches of similar processes at hadron colliders, see {\it e.g.} Refs.~\cite{Alioli:2020kez, Ellis:2022zdw, Ellis:2023ucy}.}   
In general, the high precision measurements that can be achieved at future lepton colliders makes the EFT interpretation more robust, and the interference terms between SM and the dim-8 contribution dominate over the dim-8 square terms.
It was also pointed out in \cite{Gu:2020ldn} that the leading dim-6 contribution is strongly suppressed in the process $\eeaa$ (or $\mumuaa$), so the measurement of its cross section provides a direct probe of the positivity bounds of the corresponding dim-8 operators.  
However, none of the proposed future lepton collider projects has been approved and their timelines are still unclear at the current moment.

In recent years, both ATLAS and CMS experiments have measured exclusive photon-fusion processes, including $\gamma\gamma\to \ell^+\ell^-$, $\tau^+\tau^-$, $\gamma\gamma$, $W^+W^-$, $ZZ$, and $\pi^+ \pi^-$~\cite{ATLAS:2017sfe, ATLAS:2020mve, ATLAS:2020iwi, CMS:2022dmc, TOTEM:2021zxa, ATLAS:2022uef}.  
Crucially, these photon-fusion processes are identified by requiring at least one intact forward protons, which can be measured with the ATLAS forward proton tagger (AFP)~\cite{Adamczyk:2017378, ATL-PHYS-PUB-2017-012} or the TOTEM detector at CMS~\cite{Albrow:1753795}. 
Similar processes have also been measured and extensively studied in heavy ion collision experiments with forward nucleus taggers, and are often denoted as ultraperipheral collisions (UPC) processes~\cite{Baltz:2007kq, Klein:2020fmr, ATLAS:2020hii, ATLAS:2022srr, Ogrodnik:2023qzw, ATLAS:2022ryk}. 
The proton tagging also significantly reduces QCD background, as those would not leave the proton intact.  
In particular, the reverse of the process $\eemmaa$, $\aall$, is a very clean channel at the LHC, probes the same set of operators as $\eemmaa$ while also benefits from a suppressed dim-6 contribution.  It thus provides an alternative to the measurements at future lepton colliders.\footnote{This process is also of high theoretical interests in other aspects.  See {\it e.g.} Refs.~\cite{Demirci:2021zwz, Shao:2022stc, Shao:2023zge, Shao:2023bga}.} 
Other photon-fusion processes can also be exploited to probe dim-8 operators and positivity bounds.

In this paper, we perform a phenomenological study to demonstrate the physics potential of the $\aall$ channel in probing dim-8 operators and their positivity bounds.   
We calculate the parton-level differential cross section in $\aall$ and use the effective photon parton distribution function (PDF) in Ref.~\cite{Shao:2022cly} to obtain the full differential cross section.  By a comparison with the ATLAS $\aall$ analysis~\cite{ATLAS:2020mve}, we apply a proton tagging efficiency to our results and then scale the luminosity to obtain the projection of this measurement after the entire HL-LHC.  We find that, while the future lepton colliders still provide better reaches on the corresponding dim-8 operator coefficients, the reach of the HL-LHC is comparable and within the same order. 
However, at the LHC the reach on the new physics scale ($\Lambda_8$) is comparable to the maximum invariant mass of the lepton pairs, $m_{\ell\ell}$.  An upper cut on $m_{\ell\ell}$, while reducing the reach on $\Lambda_8$, improves the situation of EFT validity to some extent.  
On the other hand, we have checked that the dim-8 squared contributions (which are formally at the same order as the interference of SM and dim-12 operators) have relatively small effects compared with the interference term.  
We apply the same methods to the $\aqaq$ process, which probes a similar set of dim-8 operators, with leptons replaced by quarks.  Due to the large quark PDF 
(which however could be different from the conventional quark PDF for this kind of semi-dissipative process), this process has a much larger cross section, leading to a potentially much better reach on the operator coefficients, as well as a more robust EFT interpretation.  However, a detailed experimental study is needed to estimate the background of this channel before such a claim can be made.

The rest of this paper is organized as follows. In \autoref{sec:aall}, we lay out the theoretical framework of the $\aall$ processes, the treatment of the effective photon PDF before doing a collider analysis to obtain the projection of this measurement at the HL-LHC.  
The analysis on $\aqaq$ is presented in \autoref{sec:aqaq}. 
We conclude in \autoref{sec:con} with a discussion on other possible measurements of dim-8 operators with the forward proton/nucleus tagger. 
More details on the dim-8 operators in $\aall$ and their positivity bounds are provided in \autoref{app:dim8}.


\section{The $\aall$ process}
\label{sec:aall}


\subsection{The parton-level process and positivity bounds}

It was shown in Ref.~\cite{Gu:2020ldn} that the leading new physics contribution in the $\eeaa$ process comes from dim-8 operators.  This is due to several 
reasons: First, the tree-level SM amplitude has only one helicity configuration, where the two fermions (and the two photons) have opposite helicities.  One insertion of the dim-6 dipole operator generates a different helicity amplitude that does not interfere with the leading SM amplitude.  Second, two insertions of dipole operators, formally at the dim-8 level, can be safely neglected due to the strong constraints from low energy experiments~\cite{Bennett:2006fi, Bennett:2008dy, Andreev:2018ayy, Muong-2:2021ojo}.  Third, while many dim-6 operators contribute at the one-loop level, they generally contribute at the tree level to other EW processes that are measured with a similar or even better precision.  If those experimental constraints are considered, one could safely ignore the one-loop contributions in the $\eeaa$ process.  

Due to crossing symmetry, similar arguments also apply to the $\aall$ processes, 
except in this case one does not have a future lepton collider to constrain some of the dim-6 operators that could enter $\aall$ at the one-loop level.   
Nevertheless, many of these can be probed at the LHC as well.  In particular, the anomalous triple gauge coupling $\lambda_Z$ (corresponding to the operator $\mathcal{O}_{3W} = \frac{1}{3!} g \epsilon_{abc} W^{a\,\nu}_\mu
W^b_{\nu \rho} W^{c\,\rho\mu} $) can be constrained at the level of $10^{-3}$ with the diboson measurements at the HL-LHC~\cite{Baglio:2018bkm, Azzi:2019yne}.  Combined with the constraints from LEP, the third argument above still generally holds.  
The crossing symmetry also implies that we could directly infer from the results in Ref.~\cite{Gu:2020ldn} to obtain (for one lepton generation)
\begin{equation}
\frac{\mathrm{d} \sigma\left(
\aall \right)}{\mathrm{d}|\cos \theta|}  = \frac{e^4}{8 \pi s}\left[\frac{1+c_\theta^2}{1-c_\theta^2}+\left(a_L+a_R\right) \frac{s^2\left(1+c_\theta^2\right)}{8 e^2 v^4}\right] \,, \label{eq:dsigma}
\end{equation}
where $c_\theta \equiv \!\cos\theta$, $\theta$ is the production polar angle, $s$ is the Mandelstam variable, $a_L$ ($a_R$) is the coefficients of the $\gamma\gamma \bar{e}_L e_L$ ($\gamma\gamma \bar{e}_R e_R$) dim-8 contact interaction.  The definitions of $a_L$ and $a_R$ and their connections with the dim-8 operator coefficients are provided in \autoref{app:dim8}.  Note that we have summed over the final state polarizations of the leptons in \autoref{eq:dsigma}.  In this case, the experiment is only sensitive to the combination $a_L+a_R$. 
The positivity bounds are simply 
\begin{equation}
a_L \geq 0 \,, \hspace{1cm}  a_R \geq 0 \,,
\end{equation}
which implies that 
\begin{equation}
\frac{d \sigma}{d|\!\cos\theta|} (\aall) \ge  \frac{d  \sigma_{\rm SM} }{d|\!\cos\theta|} (\aall) \,. \label{eq:posxaa}
\end{equation}
In principle, the positivity bound applies to all fermion-pair final states, including also $\aaqq$.  We focus on the leptonic final states, $e^+e^-$ and $\mu^+\mu^-$, to avoid a potential large QCD background.  It should be noted that for $\gamma\gamma \to \tau^+\tau^-$, if the final state tau polarization can be measured, one could in principle discriminate the effects of $a_L$ and $a_R$ and separately constrain them.  We leave a detailed analysis of the $\gamma\gamma \to \tau^+\tau^-$ channel to future studies.


\subsection{The $\pllp$ process with effective photon PDF}
\label{sec:pdf}

An automated generation tool for exclusive photon fusion processes was developed in Ref.~\cite{Shao:2022cly}.  It implements the Equivalent Photon Approximation (EPA), and calculates the cross sections using photon flux derived from electric dipole and charge form factors. 
Ref.~\cite{Shao:2022cly} essentially provides an effective photon PDF\footnote{Strictly speaking, the term ``PDF'' is inaccurate since the proton remains intact.} for proton and nuclear collisions at any energy.   Using the results in Ref.~\cite{Shao:2022cly}, we convolute the parton-level cross section of \autoref{eq:dsigma} to the proton/nucleus level with two general nuclei $A$ and $B$, 
\begin{equation}
\frac{d \sigma\left(A B \rightarrow A\left(\gamma \gamma \rightarrow l^{+} l^{-}\right) B
\right)}{d|\cos \theta|}=\int \frac{d E_{\gamma_1}}{E_{\gamma_1}} \frac{d E_{\gamma_2}}{E_{\gamma_2}} \frac{\mathrm{d}^2 N_{\gamma_1 / \mathrm{Z}_1, \gamma_2 / \mathrm{Z}_2}^{(\mathrm{AB})}}{\mathrm{d} E_{\gamma_1} \mathrm{~d} E_{\gamma_2}} \frac{d \sigma\left(\gamma \gamma \rightarrow l^{+} l^{-}\right)}{d|\cos \theta|} \,,   \label{eq:dsigmapp}
\end{equation}
where
\begin{equation}
\begin{gathered}
\frac{\mathrm{d}^2 N_{\gamma_1 / \mathrm{Z}_1, \gamma_2 / \mathrm{Z}_2}^{(\mathrm{AB})}}{\mathrm{d} E_{\gamma_1} \mathrm{~d} E_{\gamma_2}}=\int \mathrm{d}^2 \boldsymbol{b}_1 \mathrm{~d}^2 \boldsymbol{b}_2 P_{\text {no inel }}\left(\left|\boldsymbol{b}_1-\boldsymbol{b}_2\right|\right) N_{\gamma_1 / \mathrm{Z}_1}\left(E_{\gamma_1}, \boldsymbol{b}_1\right) N_{\gamma_2 / \mathrm{Z}_2}\left(E_{\gamma_2}, \boldsymbol{b}_2\right) \\
\theta\left(b_1-\epsilon R_{\mathrm{A}}\right) \theta\left(b_2-\epsilon R_{\mathrm{B}}\right) \,.
\end{gathered}
\end{equation}
In the expression, the step function $\theta\left(b-\epsilon R \right)$ enforces a constraint on the impact range, characterized by the impact parameter $b$, the nuclear radius $R$ as well as the positive parameter $\epsilon$. Typically, in the Electric Dipole Form Factor (EDFF) method, $\epsilon$ is set to $1$. Additionally, $P_{\text {no inel }}(b)$ denotes the probability of having no inelastic hadronic interaction. For simplicity, we take a reasonable approximation $P_{\text {no inel }}(b)=1$, implying a 100\% the survival rate. As demonstrated in Ref.~\cite{Shao:2022cly}, this approximation does not significantly affect the outcomes for p-p collisions. The term $N_{\gamma_1 / \mathrm{Z}_1}\left(E_{\gamma_1}, \boldsymbol{b}_1\right)$ describes the photon number density, derived by EDFF method:
\begin{equation}
N_{\gamma / \mathrm{Z}}^{\mathrm{EDFF}}\left(E_\gamma, b\right)=\frac{Z^2 \alpha}{\pi^2} \frac{\xi^2}{b^2}\left[K_1^2(\xi)+\frac{1}{\gamma_{\mathrm{L}}^2} K_0^2(\xi)\right] \,,
\end{equation}
where  where $Z$ is the number of protons in the nucleus, $\xi=E_\gamma b / \gamma_{\mathrm{L}}$, and $K_i$ represents modified Bessel functions. The parameter $\alpha$ is the fine structure constant, while $Z$ is the nuclear charge.

For proton collisions, one simply sets $Z=1$ and $R_A=R_B=R_{\rm proton}$.


\subsection{Collider reach}

With the full differential cross section in \autoref{eq:dsigmapp}, we are now ready to perform a collider study to obtain the precision reach on the $a_L$ and $a_R$ parameters at the HL-LHC.  
To estimate the selection efficiency, we use MadGraph5~\cite{Alwall:2014hca} with the built-in photon PDF package from Ref.~\cite{Shao:2022cly} to generate $\pllp$ events.  We then apply the main selection cuts in Ref.~\cite{ATLAS:2020mve}. First, each lepton is required to have $p_T>15\,$GeV and $\eta<2.5$  --- these cuts do not have a significant impact on the signal events, as we checked that the signal cross section only reduced by $3\%$ compared with the MadGraph default cuts.   
We then require the invariant mass of the lepton pair $m_{\ell\ell}$ to be larger than $105\,$GeV to avoid $Z$ backgrounds.  The events with lower invariant mass are also less useful for our analysis due to the large energy enhancements of dim-8 effects.   
After these basic cuts, we compare the cross section of the simulated events with the reported one in Ref.~\cite{ATLAS:2020mve}.  The ratio between them provides a simple estimation of the selection efficiency of the forward proton tagger, which we find to be around $11\%$. 
We compare the invariant mass distribution $m_{\ell\ell}$ and found a reasonable agreement between the simulation after applying the $11\%$ efficiency. 
We also note from Ref.~\cite{ATLAS:2020mve} that the background in the $\pllp$ channel is much smaller than the signal after the selection cuts, so we will simply ignore the background in our study.  

After the comparison with Ref.~\cite{ATLAS:2020mve}, we then scale the luminosity to the HL-LHC projection (two detectors, each with $3\inab$) and perform a simple $\chi^2$ analysis to extract the reach on $a_L$ and $a_R$.  
As mentioned above, a universal proton-tagging efficiency of $11\%$ is applied. 
Different from the lepton collider case which has a fixed center-of-mass energy, here we have a distribution for the center-of-mass energy (or the invariant mass of the lepton pair, $m_{\ell\ell}$), which contains crucial information.  To extract it, we perform a binned analysis, dividing $m_{\ell\ell} \in [105,1500]$\,GeV into 14 bins with a bin width of $100$ GeV (except the first bin which is 95\,GeV wide).  A $\chi^2$ is constructed for each bin based on the statistical uncertainty of the fiducial cross section, which are combined (assuming no correlation among different bins) to form the total $\chi^2$.

\begin{figure}[t]
    \centering
    \includegraphics[width=.55\textwidth]{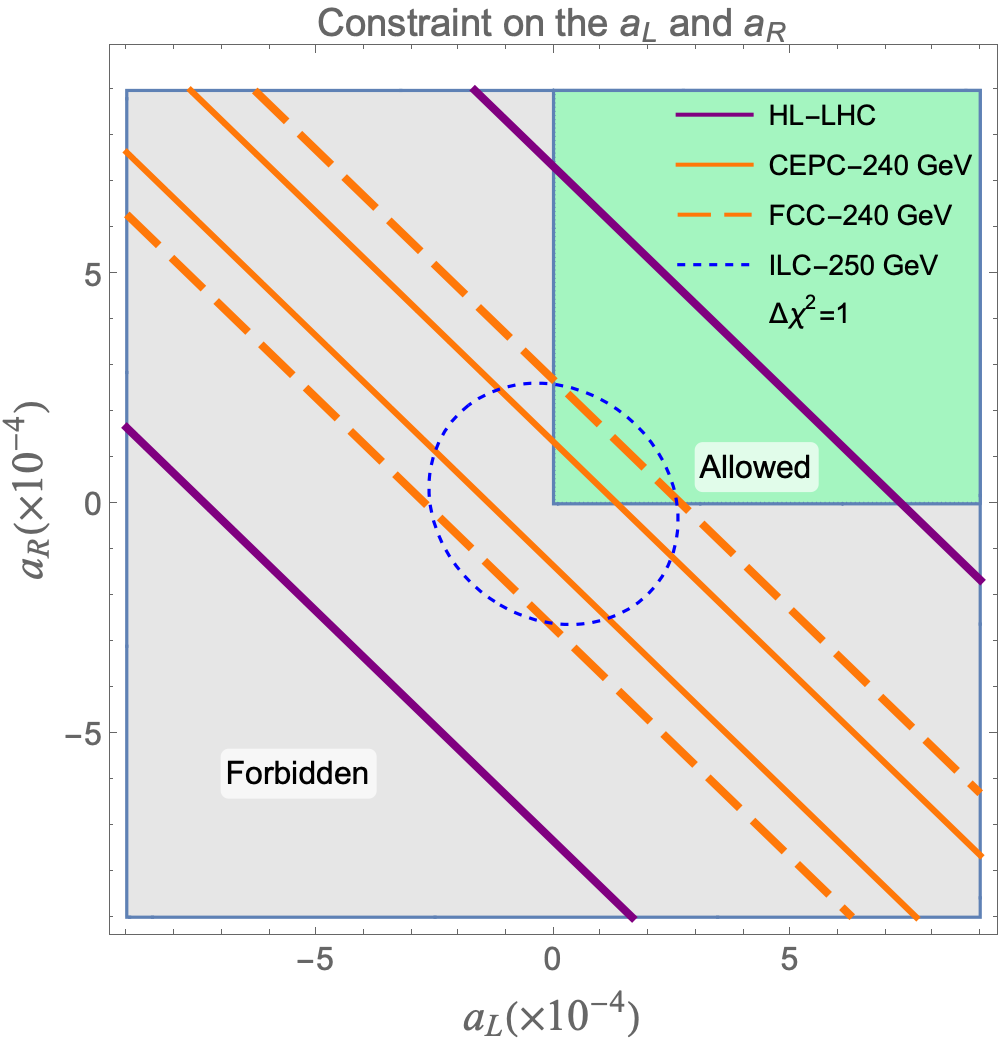}
    \caption{  The projected precision reach on the $a_L$ and $a_R$ from the $\aall$ measurement at HL-LHC and the $\eeaa$ measurements at future lepton colliders.  For the latter, the results are reproduced from Ref.~\cite{Gu:2020ldn} but with updated CEPC 240\,GeV luminosity ($20\inab$).  Only the green area is allowed by positivity bounds.  Note that without beam polarization or other means to separate different initial or final helicity states, only the combination $a_L+a_R$ can be constrained.  For the $\aall$ measurement, a maximum value of 1.5\,TeV is imposed on the invariant mass of the lepton pair, $m_{\ell\ell}$.     
    }
    \label{fig:aa1}
\end{figure}

The $\Delta \chi^2=1$ contour is shown in \autoref{fig:aa1} for the projected HL-LHC $pp$ run with a total luminosity of $2\times3\inab$.  We have imposed an upper bound of 1.5\,TeV on the maximal value of $m_{\ell\ell}$.  For comparison, we also shown the bounds from the $\eeaa$ measurement at the 240\,GeV run of CEPC~\cite{CEPCStudyGroup:2018ghi} and FCC-ee~\cite{Abada:2019zxq} and the 250\,GeV run of ILC~\cite{Bambade:2019fyw}, reproduced from Ref.~\cite{Gu:2020ldn} with updated run scenario for the CEPC~\cite{CEPCPhysicsStudyGroup:2022uwl}. 
The difference between CEPC and FCC-ee is due to their different luminosities, which is $20\inab$ for CEPC-240\,GeV and $5\inab$ for FCC-240\,GeV according to the current projections.  As pointed out in Ref.~\cite{Gu:2020ldn}, without the knowledge of the polarization (or helicity) of the initial or final state particles, only the combination $a_L+a_R$ can be constrained.  This is the case for CEPC and FCC-ee, and HL-LHC as well.  We find that the (1-sigma) precision on $a_L+a_R$ could reach around $7.3\times 10^{-4}$ at the HL-LHC, which is about a factor of 3 worse than the one of FCC-ee.

\begin{figure}
    \centering
        \includegraphics[width=.45\textwidth]{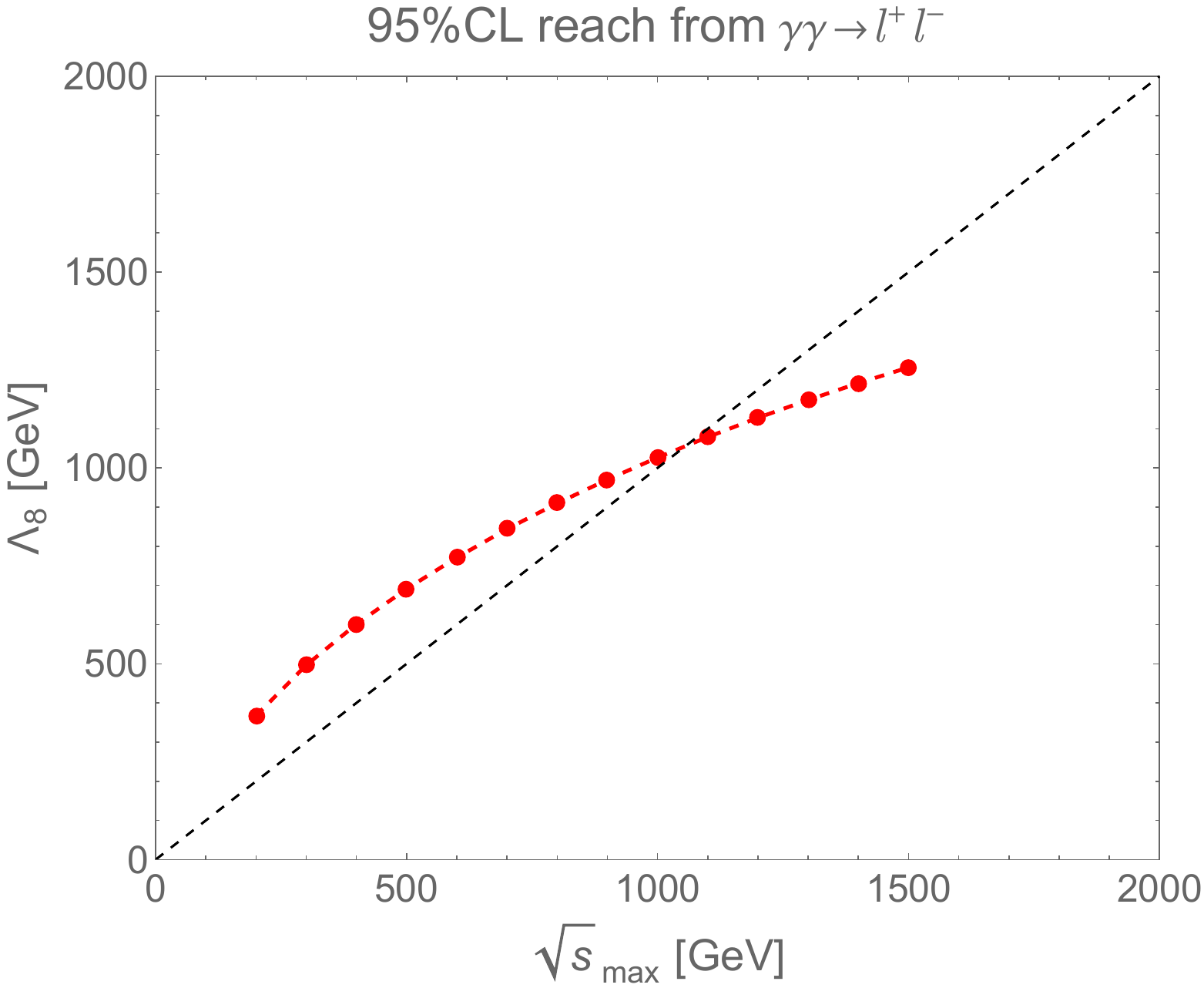}
        \includegraphics[width=.45\textwidth]{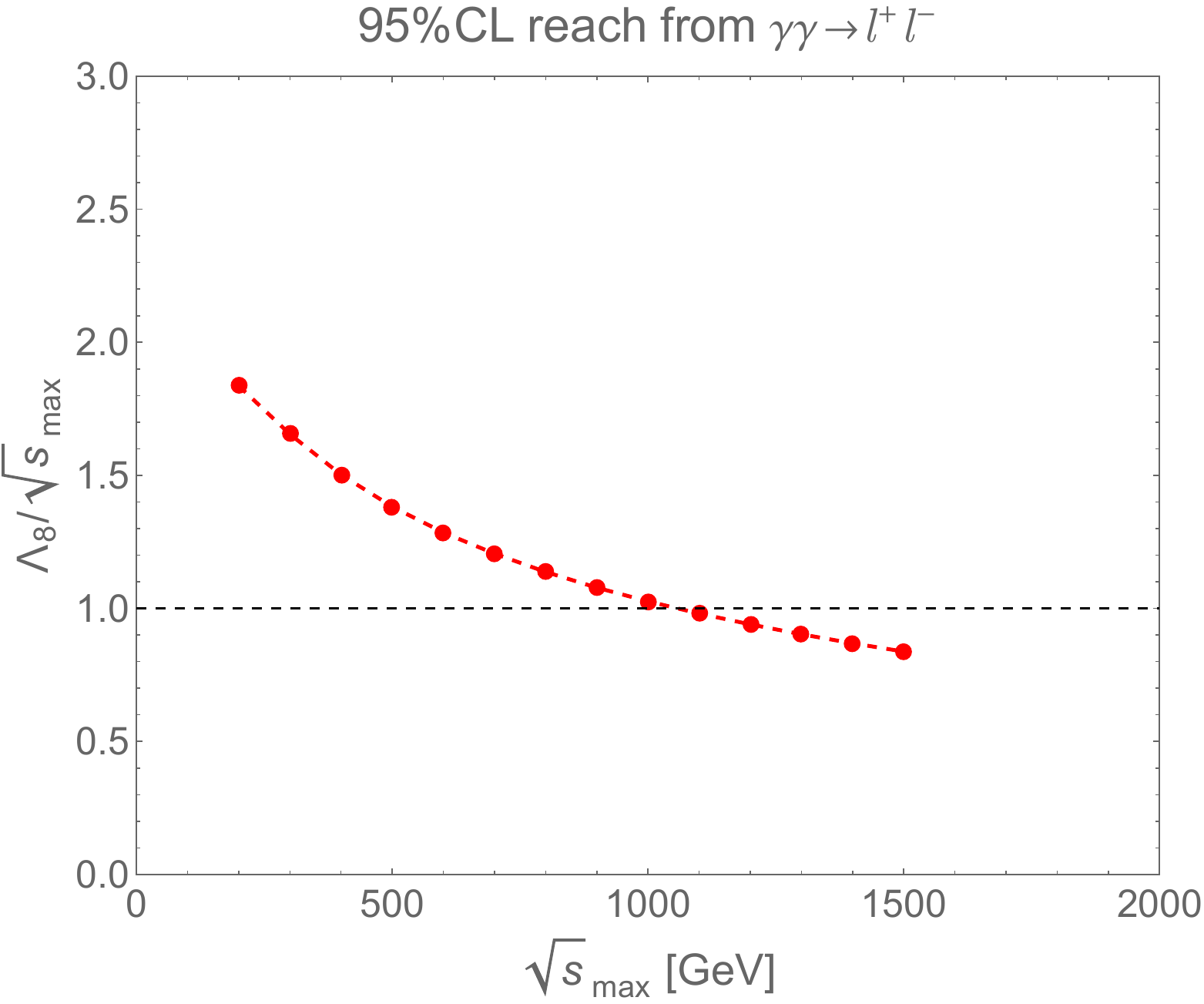}
    \caption{{\bf Left:} The 95\%\,CL reach on $\Lambda_8 \equiv v/a^{1/4}$ from the $\aall$ measurement at HL-LHC as a function of the upper bound of the lepton-pair invariant mass, $\sqrt{s}_{\rm max}$.  {\bf Right:} The same result shown in terms of $\Lambda_8/\sqrt{s}_{\rm max}$.  The dotted line corresponds to $\Lambda_8= \sqrt{s}_{\rm max}$ in both panels.    
    }
    \label{fig:lambda1}
\end{figure}

As mentioned above, the large center-of-mass energy ($m_{\ell\ell}$) at the LHC raises the question on the validity of EFT.  
One way to quantitatively address this problem is to impose an upper cut on $m_{\ell\ell}$ and compare it with the scale of new physics that can be reached from the analysis, which requires assumptions on the coupling strength~\cite{Contino:2016jqw}.
Assuming a coupling of order one, We take a na\"ive definition of the new physics scale, $\Lambda_8 \equiv v/a^{1/4}$ where $a=a_L+a_R$,\footnote{More explicitly, one could write the Wilson coefficient as $\frac{c_8}{\Lambda^4_8} \equiv \frac{a}{v^4}$, where $c_8$ is a dimensionless coupling strength.  Setting $c_8=1$ gives $\Lambda_8 = v/a^{1/4}$.  } 
and examine how the 95\% confidence-level (CL) reach on $\Lambda_8$ changes for different choices of upper bounds on $m_{\ell\ell}$ (which we denote as $\sqrt{s}_{\rm max}$). 
Our results are shown in \autoref{fig:lambda1}.  For better visualization, we have shown both $\Lambda_8$ (left panel) and $\Lambda_8/\sqrt{s}_{\rm max}$ (right panel), while the dotted line corresponds to $\Lambda_8= \sqrt{s}_{\rm max}$ in both panels.  As we can see, with our default cut $m_{\ell\ell}<1.5\,$TeV, the $\Lambda_8$ that can be reached is comparable with $\sqrt{s}_{\rm max}$.  A smaller $\sqrt{s}_{\rm max}$ improves $\Lambda_8/\sqrt{s}_{\rm max}$, but also worsen the reach on $\Lambda_8$.

\begin{figure}
    \centering
    \includegraphics[width=.95\textwidth]{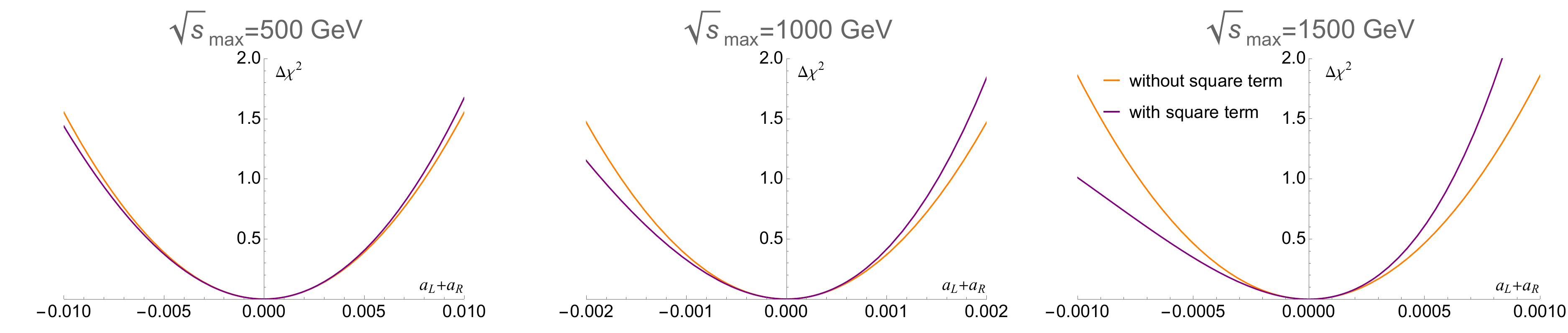}
    \caption{ The $\Delta \chi^2$ as a function of $a=a_L+a_R$ from the $\aall$ measurement at HL-LHC, 
    without (orange) or with (purple) the dim-8 squared terms.  The results are shown for 3 choices of $\sqrt{s}_{\rm max}$ (maximum invariant mass of the lepton pair), $500\,$GeV (left), $1000\,$GeV (middle) and $1500\,$GeV (right).      
    }
    \label{fig:squ1}
\end{figure}

So far, our analysis only considers dim-8 contributions via their interference with the SM. The impacts of the dim-8 squared terms are shown in \autoref{fig:squ1}.  
The square terms may also be considered as an theory error from missing higher order contributions~\cite{Alte:2017pme}, but one should note that the dim-8 squared terms are of the order $\Lambda^{-8}$, while the dim-10 operators are at $\Lambda^{-6}$.  The results are shown for 3 choices of $\sqrt{s}_{\rm max}$, $500\,$GeV (left panel), $1000\,$GeV (middle panel) and $1500\,$GeV (right panel).  In each ease, we show the $\Delta \chi^2$ as a function of $a=a_L+a_R$ both without and with the dim-8 squared terms.  We see that, even for $\sqrt{s}_{\rm max}=1500\,$GeV, the impact of the dim-8 square terms is relatively small, which changes the 1-sigma bound on $a$ from $\pm 7.3\times 10^{-4}$ to $[-9.9\,,\,6.2]\times10^{-4}$.

As mentioned above, the statement on the EFT validity depends on the coupling strength of the underlying theory.  If a strongly coupled theory is assumed ({\it e.g.}, as in Ref.~\cite{Bellazzini:2017bkb}), one typically has $c_8 \gg 1$, in which case the reach on $\Lambda_8$ is significantly larger than what we have obtained assuming $c_8=1$.  Correspondingly, the dim-12 contribution ($\sim c_{12}/\Lambda$) is suppressed compared with the dim-8 squared contribution ($\sim c^2_8/\Lambda$) since $c^2_8 \gg c_{12}$.  In such cases, a more robust interpretation of the experimental result as a test of positivity bounds can be made.   


We repeat the same analysis for the Pb-Pb collision experiment at HL-LHC.  The cross section of the UPC process has a large enhancement proportional to $Z^4$, as $Z$ is also the electric charge of the nucleus.  However, the heavy ion collision runs generally have much smaller luminosities, while 
the PDF of photon also drops significantly faster at large $x$~\cite{Shao:2022cly}.  
Assuming a center-of-mass energy of $14\,$TeV and a total luminosity of $40\, {\rm nb}^{-1}$ (which roughly corresponds to the total integrated luminosity at the end of Run 4~\cite{Bruce:2722753}), even with a 100\% tagging efficiency and $P_{\text {no inel }}=1$, we find that the reach on $a_L+a_R$ is only at the $10^{-1}$ level, which is significantly worse than the projected HL-LHC $pp$ run result. The result is summarized in \autoref{tab:ion}. 

\begin{table}[h]
\centering
\begin{tabular}{lcccc}
\hline \hline
\textbf{System} & \textbf{$\sqrt{s}$} & \textbf{Luminosity} & \textbf{Radius} & \textbf{1$\sigma$ bound on $a_L+a_R$} \\
\hline
Pb-Pb & 14 TeV & \(40\, \text{nb}^{-1}\) & \(7.1\, \text{fm}\) & $\sim 0.1$ \\
\hline \hline
\end{tabular}
\caption{The result of the $\aall$ measurement in Pb-Pb collision at HL-LHC.  We assume a total luminosity of $40\, {\rm 
nb}^{-1}$, a $100\%$ signal selection efficiency and $P_{\text {no inel }}=1$.   A maximum value of $1.5\, {\rm TeV}$ is imposed on the invariant mass of the lepton pair. 
}
\label{tab:ion}
\end{table}
%


\section{The $\aqaq$ process}
\label{sec:aqaq}

Similar to the $\aall$ process, the hadronic process, $\aaqq$, can also be measured and is sensitive to a set of dim-8 operators involving quarks, which are subject to positivity bounds as well.  However, it may suffer from a large QCD background and it is not clear whether a meaningful sensitivity could be reached (except maybe the $b\bar{b}$ case for which b-tagging could be exploited).  On the other hand, the process $\aqaq$ is sensitive to the same set of operators, and may provide better sensitivities. To the best of our knowledge, it has not been measured so far.  This process has two advantages: first, the final state photon could be used to reduce QCD backgrounds; second, the initial state quark has a larger PDF than photon, so the cross section of this process is larger, especially at high energies.  A proper estimation of backgrounds of this process is beyond the scope of this paper.  As a proof of principle, here we perform an ideal signal-only analysis to estimate the best-possible reach of this channel.  
We also note that $\aqaq$ is also related to $q\bar{q} \to \gamma\gamma$ by crossing, while the latter could benefit from an even larger event rate.  The measurement of the $q\bar{q} \to \gamma\gamma$ process provides important and complementary probes to the dim-8 operators and their positivity bounds~\cite{zhenaa}.

The $\aqaq$ amplitude is related to the $\aaqq$ one by a $s \leftrightarrow t$ crossing,  The parton level differential cross section is given by
\begin{equation}
\frac{\mathrm{d} \sigma\left(\aqaq \right)}{\mathrm{d}\cos \theta}  =\frac{Q^4 e^4}{16 \pi s}\left[\frac{2}{1+c_\theta}-\left(a_L+a_R\right) \frac{s^2}{2 Q^2 e^2 v^4}\right] \,,  \label{eq:sigmaqa}
\end{equation}
where $Q$ is the charge of $q$.
Here, from the proton tagger we know which side the photon comes from, and the two final states are also distinguishable, so the sign of $\cos\theta$ is measurable. 
Note that, due to crossing, now the interference term is destructive, and the positivity bounds $a_L,\,a_R\ge 0$ implies that 
\begin{equation}
\frac{d \sigma}{d\cos\theta} (\aqaq) \le  \frac{d  \sigma_{\rm SM} }{d \cos\theta} (\aqaq) \,. \label{eq:posxaa2}
\end{equation}

We convolute \autoref{eq:sigmaqa} with the ``photon PDF'' in \autoref{sec:pdf} and the quark PDF to obtain the hadron-level cross section.  
Again, here the ``photon PDF'' is an inaccurate term since the proton needs to be intact for the event to be selected by the forward proton tagger.  The proton that produces a quark, on the other hand, needs to be explicitly broken.  It is unclear what the rate of such an ``semi-dissipative'' process is, and whether the conventional quark PDF would provide a reasonable approximation.\footnote{The $\aqaq$ process can also be produced with the real photon PDF~\cite{Manohar:2016nzj}, but the signal selection can be extremely challenging without the proton tagger.  On the other hand, there is a diffractive quark parton distribution function~\cite{Hatta:2022lzj} that leaves the proton intact, but it is unclear whether it is relevant at high energy.}  
Needless to say, the precise estimation of the SM rate of this process is crucial for the study of the dim-8 contribution, but is beyond the scope of this paper.    
For simplicity, here we use the MMHT quark PDF~\cite{harland2015parton}. 
We also assume that the proton tagging efficiency to be 5.5\%, half of the $11\%$ of the $\aall$ process.\footnote{This assumes the proton tagger is only on one-side of the detector, and the tagging efficiency is independent of the process.  If there are proton taggers on both sides, one would instead conclude from the comparison with Ref.~\cite{ATLAS:2020mve} that each tagger has an efficiency of $1-\sqrt{1-11\%}\approx5.7\%$, which is the tagging efficiency of the $\aqaq$ process.  Numerically this makes very little difference.}

\begin{figure}[t]
    \centering
        \includegraphics[width=.45\textwidth]{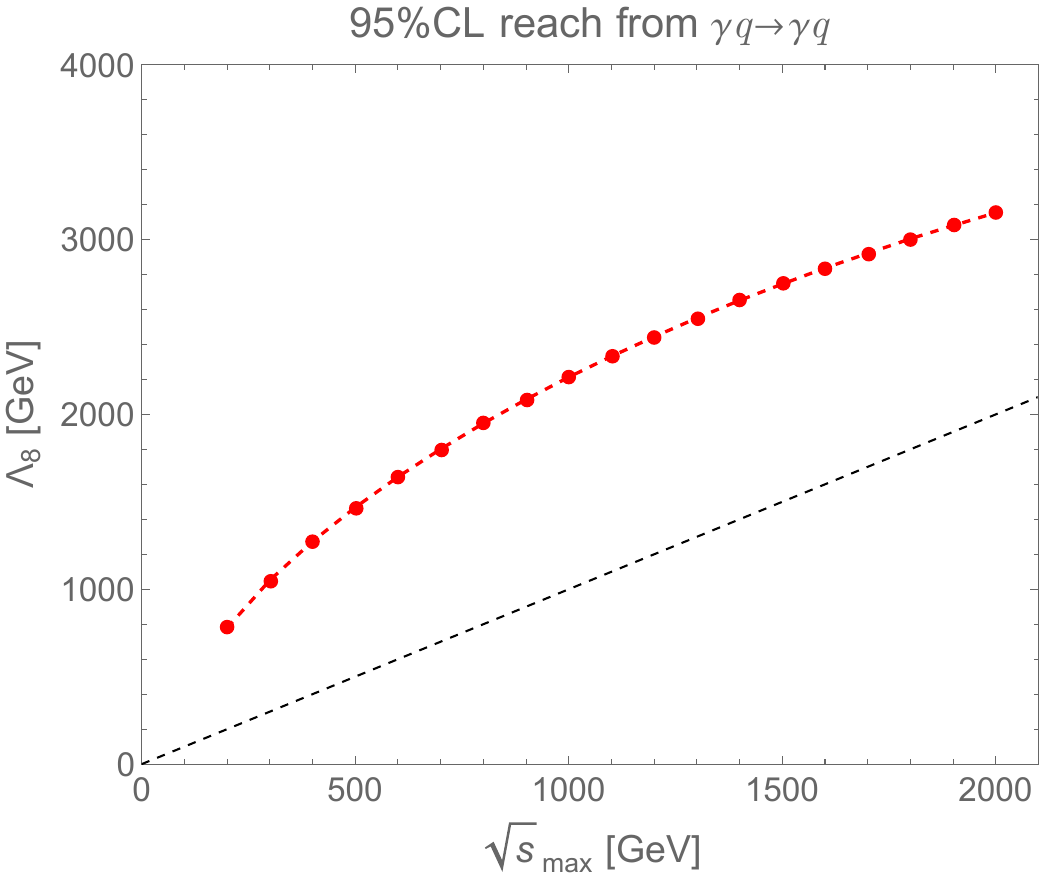}
        \includegraphics[width=.45\textwidth]{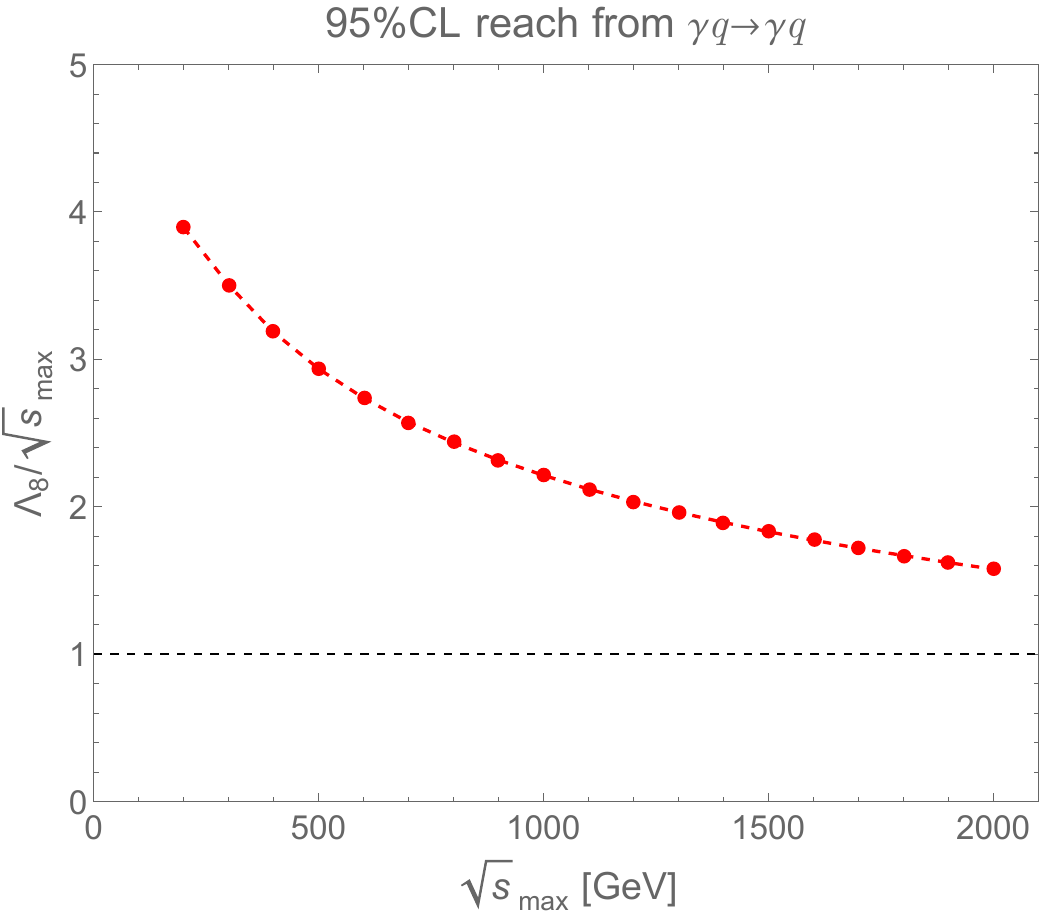}
    \caption{ Same as \autoref{fig:lambda1} but for $\aqaq$.  $a_L$ and $a_R$ are assumed to be universal for all quark flavors.  
    }
    \label{fig:qaqa}
\end{figure}
It is generally difficult to tag the final state quark flavor (except for the $b$ quark).  Here we make a simple assumption that $a=a_L+a_R$ are universal for all quark flavors, and use the inclusive semi-dissipative process $pp \to p\gamma j+X$ to constrain $a$.  We assume statistical uncertainties only and impose a upper bound ($\sqrt{s}_{\rm max}$) of 1.5\,TeV on the $\gamma$-jet pair invariant mass.  After performing a $\chi^2$ analysis with binned invariant mass distribution (similar to the one of $\aall$),   
we obtain a 1-sigma bound of $\pm 1.8 \times10^{-5}$  
for the $a$ parameter of quark operators.  The results are also shown in \autoref{fig:qaqa} in terms of the 95\%\,CL reach on $\Lambda_8 \equiv v/a^{1/4}$ as a function of $\sqrt{s}_{\rm max}$.  Compared with the same result for $\aall$ in \autoref{fig:lambda1}, we see that a significantly larger $\Lambda_8$ can be reached for the $\aqaq$ channel, which is in the range of several TeVs.  
This also improves the robustness of the EFT validity, as a large value ($>2$) of $\Lambda_8/\sqrt{s}_{\rm max}$ can be achieved without a strict choice of $\sqrt{s}_{\rm max}$.      
It is also possible to pick out the $\abab$ process with b-tagging and probe the positivity bounds of the operator involving bottom quarks.  Assuming a 50\% b-tagging rate, we obtain a one-sigma bound of $\pm 2.0 \times 10^{-4}$ 
for the $a$ parameter of the bottom-quark operators.

We note again that the result of the $\aqaq$ channel could receive a significant impact from potential backgrounds.  With large QCD background, even a tiny mistagging rate (that a $pp \to \gamma j$ event is somehow mistagged by the forward proton tagger) could result in an overwhelming amount of backgrounds.  
An experimental study for background estimation is thus crucial for this channel. 
In addition, the process $\gamma g \to \gamma g$, while loop-suppressed in the SM, also contributes to the same channel.  Due to its different helicity structure, an angular distribution analysis will be very useful in removing this background or distinguishing the dim-8 contributions in it from the ones in $\aqaq$. 
It should also be noted that the dim-6 squared contribution from quark dipoles also contribute to the $\aqaq$ processes, and unlike $\aall$, the quark dipole operators are much less constrained.  Since their contributions to the amplitude scales as $v^2E^2$, it is possible to distinguish their effects from the ones of $a_{L,R}$ by their different energy dependence.  
We refer the readers to Ref.~\cite{zhenaa} for a more detailed analysis of the impacts of the gluon process and the dim-6-squared contributions.


\section{Conclusion}
\label{sec:con}

In this paper, we have performed a simple collider analysis of the $\aall$ process to estimate projected reaches on the corresponding dim-8 operator coefficients at the HL-LHC.  The dim-8 operator coefficients are subject to positivity bounds, making them of high theoretical interests.  For $pp$ collisions, we found a reach of $\sim 10^{-4}$ in the effective coupling combination $a_L+a_R$ of the contact $\gamma\gamma\ell^+\ell^-$ interactions,  which na\"ively (assuming couplings of order one) translates to a 95\%\,CL reach on the new physics scale $\Lambda_8$ of $\sim 1$\,TeV.  The reaches are comparable to (but slightly worse than) the ones at future lepton colliders.  Furthermore, having fermions in the final state means that it is easier to probe operators of different flavors, while for lepton colliders the initial state fermions are fixed by the collider beams.  
For heavy ion collisions, a much lower reach was found despite an enhancement from the large nucleus electric charge.  This is due to both a smaller luminosity and a smaller parton-level center-of-mass energy. 
Careful treatments are needed to ensure the validity of the EFT expansion, at least to a reasonable level.  For $\aall$, the new physic scale that can be reached turns out to be comparable with the maximum center-of-mass energy of the lepton pair.  On the other hand, we found the dim-8 squared contribution to have a relatively small impact.  Overall, the $\eemmaa$ measurements at future lepton colliders still have advantages in a more robust EFT interpretation, which is important for testing positivity bounds.

We also performed a similar analysis for $\aqaq$ in $pp$ collision with the simple assumption of a conventional quark PDF and no backgrounds.  
The reach is significantly better, with precision of $\sim 10^{-5}$ for $a_L+a_R$ (assuming the operator coefficients are universal for all quark flavors) and a reach on $\Lambda_8$ of about $2\sim 4$\,TeV. 
For $\aqaq$, we also found that a sufficiently large ratio ($\gtrsim 2$) between the new physic scale that can be reached (assuming order one couplings) and the maximum final-state center-of-mass energy can be achieved without large penalties on the reach.  With b-tagging, one is also able to pick out and probe the b-quark related operators with a reasonable sensitivity.  Precise estimations for the signal rates as well as background estimation is crucial for a more realistic projection for the reach of the $\aqaq$ channel.  Our study then provides additional motivations for such estimations to be done in the future.

With the forward proton tagger it is possible to explore a wider class of processes that are sensitive to other dim-8 operators.  Another process of interest is $\gamma\gamma \to WW/ZZ$, with its result usually interpreted in terms of bounds on dim-6 and dim-8 anomalous quartic gauge couplings. The analysis by the CMS and TOTEM collaborations~\cite{CMS:2022dmc} reported the reach on several dim-6 and dim-8 operator coefficients, some of which already at the TeV-scale level (assuming order-one couplings).  It is however challenging to separate the dim-6 and dim-8 effects, as well as the different dim-8 contributions in these channels, which makes the interpretation of positivity bounds less trivial.     
On the other hand, for the light-by-light process, $\gamma\gamma\to \gamma\gamma$, the leading new physics contributions also come as dim-8 operators.  However, the SM contribution to this process is highly suppressed, so the leading BSM contribution is in the dim-8-squared term.  The search for this process has been done in the CMS and TOTEM analyses~\cite{TOTEM:2021zxa} (see {\it e.g.} Ref.~\cite{ATLAS:2020hii} for similar measurements in nuclei collisions), and the null results were interpreted as the bounds on the coefficients of dim-8 operators $\left(F_{\mu \nu} F^{\mu \nu}\right)^2$ and $\left(F_{\mu \nu} \widetilde{F}^{\mu \nu}\right)^2$.  Since the measurements are only sensitive to the dim-8-squared term, these results unfortunately do not offer a probe on the positivity bounds.


\acknowledgments

We thank Hengne Li, Zhen Liu, Javi Serra, Ding Yu Shao, Hua-Sheng Shao, Lian-Tao Wang, and Yusheng Wu for useful discussions and valuable comments on the manuscripts.  We also thank the organizers of the UPC physics 2023 workshop where we had many useful discussions.  This work is supported by the National Natural Science Foundation of China (NSFC) under grant No.~12035008 and No.~12375091.  CS is also supported by the Hui-Chun Chin and Tsung-Dao Lee Chinese Undergraduate Research Endowment(CURE).


\appendix 
\section{Positivity bounds for dim-8 operators}
\label{app:dim8}

The $f^{+} f^{-} \gamma^{+} \gamma^{-}$ 4-point  amplitude, where $f=e_{L,R}$ and $\pm$ denotes the sign of helicities, is given by~\cite{Gu:2020ldn}
\begin{align}
\mathcal{A}\left( f^{+} f^{-} \gamma^{+} \gamma^{-} \right)_{\mathrm{SM}+\mathrm{dim-8}} = &~ 2e^2 \frac{\la 24 \ra^2}{ \la 13 \ra  \la 2 3 \ra } + \frac{a}{v^4} [13][23] \la 24 \ra^2  \nonumber\\  = &~ 2 e^2 \frac{\langle 24\rangle^2}{\langle 13\rangle\langle 23\rangle}\left(1+\frac{a}{2 e^2 v^4} t u\right),
\end{align}
where $a$ is the (dimensionless) Wilson coefficient of the contact dim-8 operator, normalized by $v^4$. To go from this amplitude to the one of interest in our paper, $\mathcal{A}\left(  \gamma^{+} \gamma^{-} f^{+} f^{-} \right)$, one simply make the exchange $1\leftrightarrow 2$ and $3\leftrightarrow 4$ for the spinor brackets.  They are also related to the elastic amplitude $\mathcal{A}\left(f^{+} \gamma^{+} f^{-} \gamma^{-}\right)$ by crossing, which is given by
\begin{equation}
\mathcal{A}\left(f^{+} \gamma^{+} f^{-} \gamma^{-}\right)_{\mathrm{SM}+\mathrm{dim}-8}=2 e^2 \frac{\langle 34\rangle^2}{\langle 12\rangle\langle 32\rangle}\left(1+\frac{a}{2 e^2 v^4} s u\right) \underset{t\to 0}\to -2e^2 \left(1 - \frac{a}{2e^2v^4}s^2 \right) \,,
\end{equation}
where the last step gives the amplitude in the forward limit, $t\to 0$.  The positivity bound, $\frac{d^2}{ds^2} \A(f^+\gamma^+ f^- \gamma^-)|_{t\to 0} \ge 0 $, implies $a\ge 0$.  We denote the $a$ parameter by $a_L$ and $a_R$ for $f=e_L$ and $f=e_R$, respectively. They are related to the dim-8 operator coefficients by
\begin{equation}
\begin{aligned}
& a_L=-2 \frac{v^4}{\Lambda^4}\left(c_W^2 c_{l^2 B^2 D}-2 s_W c_W c_{l^2 W B D}^{(2)}+s_W^2 c_{l^2 W^2 D}^{(1)}\right)\,, \\
& a_R=-2 \frac{v^4}{\Lambda^4}\left(c_W^2 c_{e^2 B^2 D}+s_W^2 c_{e^2 W^2 D}\right)\,.
\end{aligned}
\end{equation}
where the dim-8 Lagrangian is written as $\mathcal{L}_{\text{dim-8}}=\sum_i \frac{c_i}{\Lambda^4} Q_i$ and the corresponding operators $Q_i$ are given by~\cite{Murphy:2020rsh, Li:2020gnx}.
\begin{equation}
\begin{aligned}
Q_{l^2 B^2 D} & =i\left(\bar{l} \gamma^\mu \overleftrightarrow{D}^\nu l\right) B_{\mu \rho} B_\nu^{~\rho} \,,\\
Q_{l^2 W B D}^{(2)} & =i\left(\bar{l} \gamma^\mu \tau^I \overleftrightarrow{D}^\nu l\right)\left(B_{\mu \rho} W_\nu^{I \rho}+B_{\nu \rho} W_\mu^{I \rho}\right) \,,\\
Q_{l^2 W^2 D}^{(1)} & =i\left(\bar{l} \gamma^\mu \stackrel{\leftrightarrow}{D^\nu} l\right) W_{\mu \rho}^I W_\nu^{I \rho} \,,\\
Q_{e^2 B^2 D} & =i\left(\bar{e} \gamma^\mu \overleftrightarrow{D^\nu} e\right) B_{\mu \rho} B_\nu^{~\rho}  \,,\\
Q_{e^2 W^2 D} & =i\left(\bar{e} \gamma^\mu \stackrel{\leftrightarrow}{D}^\nu e\right) W_{\mu \rho}^I W_\nu^{I \rho} \,.
\end{aligned}
\end{equation}

\bibliographystyle{jhep}
\bibliography{ref}

\end{document}